\documentclass[12pt, draftclsnofoot, onecolumn]{IEEEtran}

\usepackage{cite} 
\usepackage{graphicx}
\usepackage[english]{babel}
\usepackage{amsmath}
\usepackage[linesnumbered, ruled]{algorithm2e}
\usepackage[noend]{algpseudocode}
\usepackage[font=footnotesize]{caption}
\usepackage{epsfig}
\usepackage{psfrag}
\usepackage{times}
\usepackage{subfig}
\usepackage{balance}
\usepackage{xfrac}    
\usepackage{nicefrac}
\usepackage{amssymb}
\usepackage{amsfonts}
\usepackage{bm}
\usepackage[bottom]{footmisc}

\usepackage{mwe}
\usepackage{booktabs,multirow}

\usepackage{xcolor}

\newcommand{\tred}[1]{\textcolor{black}{#1}}

\SetKwRepeat{Do}{do}{while}
\SetKwInOut{Input}{Input}
\SetKwInOut{Output}{Output}

\algrenewcommand\algorithmicforall{\textbf{foreach}}
\algrenewcommand\algorithmicindent{0.8em}

\begin{document}
\setlength\abovedisplayskip{0pt}
\captionsetup[figure]{labelfont={footnotesize},labelformat={default},labelsep=period,name={Fig.}}

\renewcommand{\textfraction}{0}
\title{Throughput Maximization of Mixed FSO/RF UAV-aided Mobile Relaying with a Buffer}
\author{Ju-Hyung Lee,~\IEEEmembership{Student Member,~IEEE,}
		Ki-Hong Park,~\IEEEmembership{Senior Member,~IEEE,}
		Young-Chai Ko,~\IEEEmembership{Senior Member,~IEEE,}
        and~Mohamed-Slim Alouini,~\IEEEmembership{Fellow,~IEEE}

\thanks{
This work was presented in part at the IEEE International Conference on Communications, Shanghai, China, May 2019.

J.-H. Lee, and Y.-C. Ko are with the School of Electrical and Computer Engineering, Korea University, Seoul, Korea (Email: leejuhyung@korea.ac.kr; koyc@korea.ac.kr)   

K.-H. Park, and M.-S. Alouini are with the Electrical Engineering Program, Computer, Electrical, Mathematical Sciences and Engineering Division, King Abdullah University of Science and Technology (KAUST), Thuwal, Makkah Province, Kingdom of Saudi Arabia (Email: kihong.park@kaust.edu.sa; slim.alouini@kaust.edu.sa).

}


}	
\date{}
\maketitle \thispagestyle{empty}

\begin{abstract}
In this paper, we investigate a mobile relaying system assisted by an unmanned aerial vehicle (UAV) with a finite size of the buffer.
\tred{
Under the buffer size limit and delay constraints at the UAV relay, we consider a dual-hop mixed free-space optical/radio frequency (FSO/RF) relaying system (i.e., the source-to-relay and relay-to-destination links employ FSO and RF links, respectively).}
Taking an imbalance in the transmission rate between RF and FSO links into consideration, we address the trajectory design of the UAV relay node to obtain the maximum data throughput at the ground user terminal.
Especially, we classify two relaying transmission schemes according to the delay requirements, i.e., \textit{i)} delay-limited transmission  and \textit{ii)} delay-tolerant transmission.
Accordingly, we propose an iterative algorithm to effectively obtain the locally optimal solution to our throughput optimization problems and further present the complexity analysis of this algorithm.
Through this algorithm, we present the resulting trajectories over the atmospheric condition, the buffer size, and the delay requirement.
In addition, we show the optimum buffer size and the throughput-delay tradeoff for a given system.
The numerical results validate that the proposed buffer-aided and delay-considered mobile relaying scheme obtains 223.33\% throughput gain compared to the conventional static relaying scheme. 
\end{abstract}


\section{Introduction} \label{Introduction}

\begin{figure}[t]
    \centering
    \includegraphics[width=0.6\textwidth]{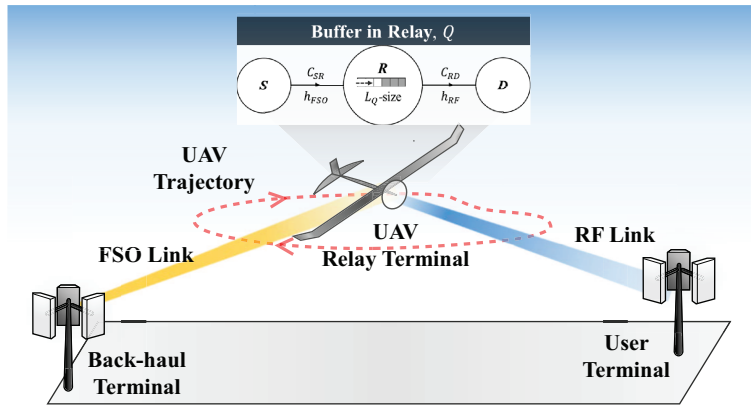}
    \caption{\tred{Illustration of a UAV-assisted mobile relaying with dual-hop mixed FSO/RF communication.}}
    \label{Fig_Intro1}
    \vspace{-1.0em}
\end{figure}

As becoming broader to utilize unmanned aerial vehicles (UAVs) for diverse applications, UAV-aided mobile relaying systems have been in the limelight \cite{System_1}.
Motivated by the attention to the unmanned flying platforms, an UAV-enabled mobile relaying framework has been proposed to transport the backhaul/fronthaul traffic between the access and core networks \cite{Intro_22, Intro_1, Intro_0}. 
In comparison to the conventional static relaying system, the mobile relaying has some key advantages. 
\tred{
Indeed, cost-effectiveness and simple deployment let the mobile relaying systems particularly suitable for unexpected events (e.g., emergency response, disaster recovery, military operation, etc.).
} 
Besides, its high maneuverability provides new opportunities for delay-tolerant applications (such as periodic sensing, massive data uploading/downloading), and performance improvements through the dynamic re-location to exploit the better communication condition.
In particular, the authors in \cite{Body_3} identified that the mobile relaying offers a new degree of freedom (DoF) for performance enhancement compared with conventional static relaying, via careful relay trajectory design.
Based on the result of \cite{Body_3}, some works including \cite{Intro_21, System_2, Intro_20} have focused on utilizing mobile relaying system in various applications.  

\tred{
\subsection{Related Works and Challenges} 
In beyond fifth-generation (B5G) with a presence of ultra-dense heterogeneous small cells, wireless backhaul/fronthaul networks requires to comply with ultra-high network throughput.
In this regard, several industry project \cite{Intro_26} and academe \cite{Intro_25, Intro_10,Intro_23, GLOBECOM} have paid attention to this challenge and researched a free space optical communication (FSO) on the mobile relaying system as a promising solution.
Particularly, by taking some advantage of FSO (e.g., unlicensed broad spectrum, immunity to electromagnetic interference, and security), Facebook had launched project \textit{Aquila} which employs FSO communication for the source to UAV-assisted relay link and UAV-to-UAV link, to support the high throughput in the air-to-ground (A2G) and air-to-air channel.
Motivated by the potential of FSO-assisted mobile relaying, our previous works of \cite{Intro_23, GLOBECOM} investigated the point-to-point (PtP) FSO link for UAV-to-ground user terminal (i.e., single-hop communication).}

\tred{
Besides, extensive works in academia have developed to exploit the benefits of FSO-based communication and also have accordingly adopted a mixed FSO/RF system to utilize both RF and FSO links.
Particularly, to consider the dual-hop system (e.g., relaying system), the mixed FSO/RF communication has been investigated \cite{Intro_2, Intro_3,Intro_4,Intro_5}.
In \cite{Intro_2}, the dual-hop mixed FSO/RF backhauling scenario using networked flying platforms (NFPs) has been proposed. 
The efficacy of the network scenario has been shown in terms of link budget and achievable data rate for only the single FSO link under different weather conditions.
Furthermore, in \cite{Intro_3,Intro_4}, the performance analysis of mixed FSO/RF systems following the general turbulent fading model has been provided.
As identified in these works, the attenuation condition is a major challenge in dual-hop mixed FSO/RF communication, especially for A2G link.
In addition, the rate imbalance problem, which caused by using other types of links, is another major concern.}


\tred{
To address the practical issues in the asymmetric dual-hop relay system (e.g., dual-hop relay transmission with mixed FSO/RF link), the conventional works \cite{Intro_5,Intro_6,Intro_7} have focused on the buffer-assisted relaying.
Note that it is practical to consider the buffering in relaying, since the actual relay system includes a buffer.
In particular, \cite{Intro_5} demonstrated that buffering can enhance system performance under various mixed FSO/RF link conditions.
In addition, the authors in \cite{Intro_7} suggested the design of FSO-based buffer-aided cooperative protocols and showed that the buffering on FSO-assisted relay constitutes an additional DoF at the expense of increased delay.
Not only just buffer-related conditions, but also average delay in the buffer-aided relaying system, have been addressed \cite{Intro_17,Intro_9, Intro_11, Intro_13}.
In \cite{Intro_17}, the two buffer-aided relaying protocols are proposed, and the throughput-delay tradeoff and average delay for the buffer-aided relaying system are further analyzed. 
Moreover, to consider the practical delay-constrained application, \cite{Intro_9} studied two transmission schemes for buffer-aided relaying; delay-limited transmission (e.g., real-time) and delay-tolerant transmission.
Delay-limited traffic, for example, supports delay-sensitive services which include voice over internet protocol, video conferencing, and monitoring of critical processes (e.g., medical packet) \cite{Body_5}. 
On the other hand, delay-tolerant traffic does not carry an urgency and can be served when the reliable reception is essential, e.g., email, instant messages, or sensor data in periodic sensing \cite{Intro_18}.
For both delay-limited and delay-tolerant transmission, authors in \cite{Intro_9} analyzed a hybrid FSO/RF backhauling policy to satisfy a certain delay requirements. }

\subsection{Contributions and Organization} 
By taking into consideration the aforementioned features, e.g., mobility, buffering, and delay consideration in relaying system, we focus on the scenario of dual-hop mixed FSO/RF backhauling with the help of buffer, which can be a promising solution to the emerging wireless backbone network as discussed in \cite{Intro_2, GLOBECOM, Intro_23}.
In such a scenario, we study on the end-to-end optimization of a mixed FSO/RF-based mobile relaying system. 
In order to deal with the transmission rate imbalance between FSO and RF links\footnote{
Since FSO communication utilizes license-free narrow beams with a wide frequency spectrum (more bandwidth can be provided for FSO link than RF link, i.e., $B_{\mathrm{FSO}} \geq B_{\mathrm{RF}}$), it can offer data rates higher than the baseline alternatives (e.g., RF communication).
} according to the relay's location, we further regard both buffer and average delay constraints on the UAV relay node.
We have summarized the main contributions of this work as follows:

\begin{itemize}
\item 
Unlike \cite{Intro_2,Intro_23}, which have investigated the only single hop scenario in mixed FSO/RF backhaul links, we deeply look into the scenario depicted in Fig. \ref{Fig_Intro1}, in which dual-hop mixed FSO/RF backhauling is operated on the UAV-assisted relaying under a limited buffer constraint.
Especially, 
we solve the trajectory optimization for the throughput maximization in this system. 
To the best of our knowledge, there is no open literature to tackle the optimization of the buffer-constrained mobile relaying system in dual-hop mixed FSO/RF links.
\item
Furthermore, we consider buffer constraints to account for practical relaying scenarios and the transmission rate imbalance in a mixed FSO/RF system, which is induced by using a different type of links (i.e., FSO link for source-to-relay and RF link for relay-to-destination).
In other words, the UAV relay node is equipped with a buffer of finite queue size and can control throughput delay and UAV mobility. 
\item Moreover, we design the system by classifying two relay transmission schemes (i.e., \textit{i)} delay-tolerant transmission and \textit{ii)} delay-limited transmission) according to the average delay requirements of the network data throughput.
Specifically, we design the system where the data should be transmitted at the relay node below the average delay.
\item To tackle these non-convex trajectory optimization problems, we propose an iterative algorithm by utilizing the successive optimization method to find the locally optimal solution.
Then, the trajectories can be established by applying quadratically constrained programming (QCP).  
\item Under different conditions, e.g., visibility, buffer size, and delay limit, the simulation results for throughput maximized trajectories are drawn. Based on the buffer and delay constraints, the throughput-delay and the throughput-buffer tradeoff are presented.
Consequently, we validate the supremacy of the proposed scheme compared to conventional schemes (e.g., static relaying \cite{Intro_1} and data-ferrying relaying \cite{Num_1,Body_3}) according to the simulation and numerical results.
\end{itemize}

The remainder of this paper is organized as follows: 
In Section \ref{System Model}, the system model for the FSO and RF links and the metrics for the buffer constraint of dual-hop mixed FSO/RF network are presented. 
The throughput maximization problem for buffer-aided mobile relaying is formulated and optimized by two delay-considered transmission schemes (e.g., delay-limited transmission and delay-tolerant transmission) in Section \ref{Body}.
In Section \ref{Numerical Result}, numerical results are presented, and concluding remarks are drawn in Section \ref{Conclusion}.

\textit{Notation:}
Throughout this paper, we use the normal-face font to denote scalars, and boldface font to denote vectors.
We use $\mathbb{R}^{D\times 1}$ to represent the $D$-dimensional space of real-valued vectors.
We also use $\|\cdot\|$ to denote the $L^2$-norm (i.e., an Euclidean norm) and $\mathrm{log}(\cdot)$ to represent a natural logarithm.
The expression $O(\cdot)$ stands for describing the Big O notation.


\section{System Model} \label{System Model}

We consider a UAV-assisted mobile relaying employing a dual-hop mixed FSO/RF communication, as represented in Fig. \ref{Fig_Intro1}. 
Specifically, the UAV-assisted relay node utilizes an FSO link to receive information from a backhaul terminal and an RF link to forward information to a user terminal\footnote{
The opposite uplink scenario can also be considered, i.e., the scenario of an FSO link for forwarding information to the backhaul terminal and an RF link for receiving information from the user terminal(s). 
An example is the application of information collection.
We note that the more general cooperative system design of the mixed FSO/RF UAV-enabled mobile relaying remains to be our future work \cite{VTC}.
}. 
Based on three-dimensional Cartesian coordinates for the position of each terminal, we consider that the backhaul terminal and the user terminal are located at $\mathbf{q}_{\mathcal{S}}=[0,0,0]^T$ and $\mathbf{q}_{\mathcal{D}}=[L,0,0]^T$, respectively, while the UAV flies at a constant altitude of $H$ within a predetermined maximum speed $V_{\mathrm{max}}$ and acceleration $A_{\mathrm{max}}$ for a period $T$.
The time varying position of the UAV node can be expressed as $\mathbf{q}_{\mathcal{R}}(t)=[x_{\mathcal{R}}(t),y_{\mathcal{R}}(t),H]^T \in\mathbb{R}^{3 \times 1}$, $0\leq t \leq T$.

For ease of exposition, a discrete-time model is considered as in \cite{System_1}.  
The given finite time horizon $T$ is divided into $N$ time intervals each with step size $\delta_{t}$ (i.e., $T=N \cdot \delta_{t}$). 
The step size $\delta_{t}$ is selected to be sufficiently small so that the UAV's location can be appropriately approximated within each slot, such as $\mathbf{q}_{\mathcal{R}}[n] \triangleq \mathbf{q}_{\mathcal{R}}(n\delta_{t}) = [x_{\mathcal{R}}(n\delta_{t}),y_{\mathcal{R}}(n\delta_{t}),H]^T = [x_{\mathcal{R}}[n],y_{\mathcal{R}}[n],H]^T \in\mathbb{R}^{3 \times 1}$,  $0 \leq n \leq N+1$.
Note that $n=0$ and $n=N+1$ denote the initial and final time slots, respectively.

In the following, we express channel and transmission rate models for FSO and RF channels, respectively, and introduce a buffer constraint which characterizes the queuing system of a practical relaying with a finite size buffer.

\subsection{System Model for FSO Link}
For an FSO link, the channel gain at a link distance $l_{\mathrm{FSO}}$, based on the Beer-Lambert Law\footnote{Note that, if other attenuation factors, e.g., rain, snow, and haze, need to be considered, the optimization framework can be solved by adjusting only some parameters, e.g., $\beta$ or $k_2$.}, can be expressed as

\begin{equation}
 h_{\mathrm{FSO}}[n]=e^{-\beta \cdot l_{\mathrm{FSO}}[n]}=e^{-\beta \cdot \|\mathbf{q}_{\mathcal{R}}[n] - \mathbf{q}_{\mathcal{S}} \| }, \ \forall n,\label{channel_FSO}
\end{equation}
\tred{
where $\beta_{\mathrm{dB}}=\frac{3.91}{V}\left(\frac{\lambda_{0}}{550 \ \mathrm{[nm]}}\right)^{-p} \ \mathrm{[dB/km]}$ depends on the wavelength $\lambda_{0}$ assumed to be 1550 [nm] as an example in this paper, $V$ is the visibility in [km], and the size distribution coefficient $p$ is determined by Kim model \cite{System_3}. }
Note that $\beta = \frac{\beta_{\mathrm{dB}} \cdot \log{10} }{10^4} \ \mathrm{[m^{-1}]}$. 

While the capacity of the FSO channel has not been known in a closed-form, the capacity bounds of FSO have been proposed in several papers.
To describe the data rate of an FSO link between source and relay, we use the lower bound of FSO capacity in intensity channel introduced in \cite{System_4}.
The average electrical SNR (ASNR) is given by $\gamma_{\mathrm{FSO}}^2=\frac{\varepsilon^2}{\sigma_{\mathrm{FSO}}^2}$, where $\varepsilon$ and $\sigma_{\mathrm{FSO}}^2$ are the average optical power and noise variance for FSO, respectively. 
The parameter, $k_1$, related to ASNR  and parameter, $k_2$, related to attenuation condition are formulated, respectively, as 

\begin{eqnarray}
&& k_1 = \begin{cases}
\frac{e^{2\alpha_{0}\mu^{*}}}{2\pi e}\left( \frac{1-e^{-\mu^{*}} }{\mu^{*}}\right)^2 \frac{\gamma_{\mathrm{FSO}}^2}{\alpha_{0}^2}  
& \text{if } 0\!<\!\alpha_{0}\!<\!\frac{1}{2} \\
\frac{\gamma_{\mathrm{FSO}}^2}{2\pi e \alpha_{0}^2}  
& \text{if } \frac{1}{2}\!<\!\alpha_{0}\!<\!1
       \end{cases} ,  \label{k1} \\
&& k_2 = 2\beta .  \label{k2} 
\end{eqnarray}
\tred{
Note that $\mu$ is the free parameter that indicates the solution to the equation $\alpha_{0}=\frac{1}{\mu^{*}}-\frac{e^{- \mu^{*}}}{(1-e^{-\mu^{*}})}$ in which the average-to-peak ratio (APR) is set to $\alpha_{0}=\frac{\varepsilon}{\lambda_{0}}$ where $\lambda_{0}$ is the peak optical power. 
}

The FSO link achievable rate for the time slot $n$ in bits/second (bps) is given by the channel gain for FSO link in \eqref{channel_FSO}, the $k_1$ and $k_2$ in \eqref{k1} and \eqref{k2}, the bandwidth in hertz (Hz) for FSO link $B_{\mathrm{FSO}}$, and the received ASNR $\bar{\gamma}_{\mathrm{FSO}}^2=h_{\mathrm{FSO}}^2\cdot\gamma_{\mathrm{FSO}}^2$ as follows: 

\begin{eqnarray}
R_{\mathrm{FSO}}[n] = \frac{B_{\mathrm{FSO}}}{2 \mathrm{log}2} \cdot \log\left( 1+k_1e^{-k_2 \cdot \|\mathbf{q}_{\mathcal{R}}[n] - \mathbf{q}_{\mathcal{S}} \|} \right) \ \mathrm{[bps]},  && \nonumber \\
\forall n. && 
\label{Rate_FSO}
\end{eqnarray}
\tred{
Note that we assume that precise pointing, acquisition, and tracking (PAT) algorithm may compensate for the pointing error.}

\subsection{System Model for RF Link}
For an RF link between the UAV-aided relay and the user terminal, the channel gain $h_{\mathrm{RF}}$ can be expressed as

\begin{equation}
 h_{\mathrm{RF}}[n]= \sqrt{\tau_{\mathrm{RF}}[n]} \cdot \Tilde{h}_{\mathrm{RF}}[n], \ \ \forall n, \label{channel_RF}
\end{equation}
where $\tau_{\mathrm{RF}}[n]$ and $\Tilde{h}_{\mathrm{RF}}[n]$ account for the effect of large-scale fading (e.g., path loss and shadowing) and the effect of small-scale fading with $\mathbb{E}\lbrace |\Tilde{h}_{\mathrm{RF}}[n]|^2 \rbrace = 1$, respectively.
Furthermore, for UAV-ground RF link, the large-scale attenuation is usually modeled with the probabilities of LoS and non-LoS (NLoS) links\footnote{
Note that, due to the shadowing effect and the reflection of signals from obstacles, different large-scale attenuation models need to be considered for LoS and NLoS links.
}.  
As in \cite{System_11}, $\tau_{\mathrm{RF}}[n]$ is expressed as 

\begin{equation}
  \tau_{\mathrm{RF}}[n]=\begin{cases}
    \beta_{0}l_{\mathrm{RF}}^{-\Tilde{\alpha}}[n], & \text{LoS link},\\
    \kappa \beta_{0}l_{\mathrm{RF}}^{-\Tilde{\alpha}}[n], & \text{NLoS link},\\
  \end{cases}
\end{equation}
where $\beta_{0}$ represents the received power at the reference distance $d_0=1$ [m], $l_{\mathrm{RF}}$ denotes a link distance between $\mathcal{R}$ and $\mathcal{D}$, $\Tilde{\alpha}$ is the path loss exponent\footnote{$\Tilde{\alpha}$ represents the more general path loss exponent $\Tilde{\alpha}\geq2$, rather than the special case of $\Tilde{\alpha}=2$ which only considers free-space condition.}, and $\kappa$ is an additional attenuation factor due to the NLoS link.
As a result, $h_{\mathrm{RF}}[n]$ is a random variable with the random occurrence of LoS and NLoS as well as the random small-scale fading. 
Accordingly, the expected channel gain by averaging over both randomness is given by \cite{System_12},

\begin{equation}
\mathbb{E}\lbrace |h_{\mathrm{RF}}[n]|^2 \rbrace = \hat{P}_{\mathrm{LoS}}[n] \beta_{0}l_{\mathrm{RF}}^{-\Tilde{\alpha}}[n], \ \forall n,
\end{equation}
where $\hat{P}_{\mathrm{LoS}}[n]=P_{\mathrm{LoS}}[n] + (1-P_{\mathrm{LoS}}[n])\kappa$, and $P_{\mathrm{LoS}}[n] = \frac{1}{1+C \cdot \mathrm{exp}(-D [ \theta[n] - C ])}$ is the LoS probability between UAV and the user terminal in which $C$ and $D$ are the parameters depending on the propagation condition, and $\theta[n]=\frac{180}{\pi}\mathrm{sin}^{-1}(\nicefrac{H}{l_{\mathrm{RF}}[n]})$ is the elevation angle in degree.

The achievable rate in bps between UAV and the user terminal with the constant transmission power $P$ at time slot $n$ is expressed as

\begin{equation}
\hat{R}_{\mathrm{RF}}[n] = \frac{B_{\mathrm{RF}}}{\log 2} \cdot \log\left( 1+ \dfrac{P |h_{\mathrm{RF}}[n]|^2}{\sigma_{\mathrm{RF}}^2} \right) \ \mathrm{[bps]},  \  \ \forall n, \label{R_RF_0}
\end{equation}
where $B_{\mathrm{RF}}$ represents the RF bandwidth in Hz, and $\sigma_{\mathrm{RF}}^2$ is the noise variance for RF.
Note that, as the channel gain $h_{\mathrm{RF}}[n]$ is the random variable, $\hat{R}_{\mathrm{RF}}[n]$ is also a random variable. 
Using the concavity of \eqref{R_RF_0} and Jensen's inequality, we have 

\begin{eqnarray}
\mathbb{E}\lbrace \hat{R}_{\mathrm{RF}}[n] \rbrace \!\!&\!\!\leq\!\!&\!\! \displaystyle\sum_{n=1}^{N} \frac{B_{\mathrm{RF}}}{N \log 2} \cdot \log\left( 1+ \dfrac{P \mathbb{E}\lbrace |h_{\mathrm{RF}}[n]|^2 \rbrace}{\sigma_{\mathrm{RF}}^2} \right) \label{R_RF_1}
\\
\!\!&\!\!=\!\!&\!\!\displaystyle\sum_{n=1}^{N} \frac{B_{\mathrm{RF}}}{N \log 2} \cdot \log\left( 1+ \dfrac{\Tilde{\gamma}_0 \hat{P}_{\mathrm{LoS}}[n]}{(\|\mathbf{q}_{\mathcal{D}} - \mathbf{q}_{\mathcal{R}}[n] \|^2)^{\alpha}} \right), \nonumber \\
&& \label{R_RF_2}
\end{eqnarray}
where $\Tilde{\gamma}_0 \triangleq \frac{\beta_0 \cdot P}{\sigma_{\mathrm{RF}}^2}$ and $\alpha \triangleq \Tilde{\alpha}/2$.
It is found that \eqref{R_RF_2} depends on the UAV location $\mathbf{q}_{\mathcal{R}}[n]$ not only over $l_{\mathrm{RF}}[n]$, but also over $\hat{P}_{\mathrm{LoS}}[n]$. 
For this reason, it is challenging to handle \eqref{R_RF_2} directly. 
To resolve this issue, we use the homogeneous approximation of the LoS probability, i.e., $\hat{P}_{\mathrm{LoS}}[n] \simeq \bar{P}_{\mathrm{LoS}}, \forall n$, as in \cite{System_12}.
Note that $\bar{P}_{\mathrm{LoS}}$ could be the value corresponding to the most likely elevation angle or the average value based on a certain heuristic UAV trajectory.
Accordingly, \eqref{R_RF_2} can be rewritten as $\bar{R}_{\mathrm{RF}} = \displaystyle\sum_{n=1}^{N} \frac{B_{\mathrm{RF}}}{N \log 2} \cdot \log\left( 1+ \dfrac{\gamma_0}{(\|\mathbf{q}_{\mathcal{D}} - \mathbf{q}_{\mathcal{R}}[n] \|^2)^{\alpha}}  \right)$ where $\gamma_0 \triangleq \Tilde{\gamma}_0 \bar{P}_{\mathrm{LoS}}$.
Thus, the achievable rate of RF link for the time slot $n$ is expressed as \cite{System_13, System_12}

\begin{equation}
R_{\mathrm{RF}}[n] = \frac{B_{\mathrm{RF}}}{\log 2} \cdot \log\left( 1+ \dfrac{\gamma_0}{(\|\mathbf{q}_{\mathcal{D}} - \mathbf{q}_{\mathcal{R}}[n] \|^2)^{\alpha}}  \right),   \ \forall n, \label{Rate_RF_4}
\end{equation}
It is worth noting that $\bar{R}_{\mathrm{RF}}$ is the approximation of $\mathbb{E}\lbrace \hat{R}_{\mathrm{RF}}[n]\rbrace$, and $R_{\mathrm{RF}}[n]$ is the corresponding average rate expression  between $\mathcal{R}$ and $\mathcal{D}$ at any time slot $n$.

\subsection{Quality of Service (QoS) Metrics for Buffer Constraint at UAV-aided Relay}
\begin{figure}[t]
    \centering
    \includegraphics[width=0.6\textwidth]{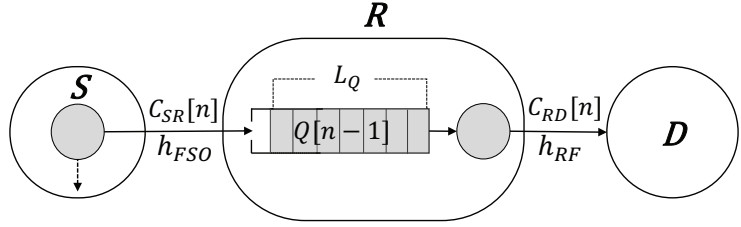}
    \caption{The considered system model for a buffer-aided relay network with a mixed FSO/RF communication. Illustration of the arrival-departure process in a buffer at relay $\mathcal{R}$ during time-step $n$.}
    \label{Fig_Intro2}
\end{figure}

As shown in Fig. \ref{Fig_Intro2}, we consider a dual-hop mixed FSO/RF network communicating between a source $\mathcal{S}$ and a destination $\mathcal{D}$ via a single UAV-enabled mobile relay node $\mathcal{R}$.
Throughout the system, it is assumed that the source transmits the message in packets via an FSO link at the rate of $\mathit{C}_{\mathcal{SR}}[n]$.
For ease of exposition, an amount of data rate instead of a packet rate is considered.
Note that analysis based on the packet rate can also be performed by dividing the bit rate with the number of bits per packet, if necessary.
For the scheduling policy at queuing node, we consider first-in-first-out (FIFO), which implies that packets are en-queued in turn, and the packets which wait longest in a buffer are de-queued first.
Thanks to the independent operation of the mixed FSO/RF transmission in parallel \cite{Intro_7}, we leverage a full-duplex relaying (FDR), which works in the decode-and-forward (DF) protocol in this system.
\tred{
Note that the error-free condition in transmission, relaying, and detection is assumed for explaining the transmission rate, in which we deal with the buffer-aided mobile relaying based on the optimization perspective instead of the performance analysis perspective. 
}
In the following, we discuss the queuing dynamics in which source and relay transmit data as in \cite{System_6, System_9}.  

\subsubsection{Source transmits}
As FSO is chosen for $\mathcal{S-R}$ link, the transmission rate of $\mathcal{S-R}$ link in time slot $n$ in bps\footnote{Since we consider the discrete-time model (e.g., $t = n \cdot \delta_{t}$), we thereby normalize the variables according to $n$ and $\delta_{t}$. Note that we assume $\delta_{t}=1$ [s] in this paper.} is given by

\begin{equation}
\mathit{C}_{\mathcal{SR}}[n] \leq R_{\mathrm{FSO}}[n], \ n=1,2, \cdots, N, \label{C_SR} 
\end{equation}
\textcolor{black}{
where $\mathit{C}_{\mathcal{SR}}[n]$ represents the data rate transmitted to queue at slot $n$.
This bound implies that the data arriving into the buffer at the source can be dropped when the buffer is full or adjusted to meet the delay requirement at the relay. 
}

\subsubsection{Relay enqueues}
Hence, the relay receives data bits at rate $\mathit{C}_{\mathcal{SR}}[n]$ from $\mathcal{S}$
and stacks them on the queue in its buffer. 
The relationship among the queue length of the buffer, transmission rate, and the drop rate above is illustrated in Fig. \ref{Fig_Intro2}.
Note that we follow the first-out scheme for admission control, as described in \cite{System_7}. 

In general, the FSO link can provide a higher data rate (in excess of tens of Gbps) than the RF link, since FSO link can use a wider bandwidth  \cite{System_8}.
\tred{
Accordingly, since the imbalance of the transmission rate between FSO and RF links can easily occur, we therefore deal with the buffer constraint in this system to avoid an buffer overflow.
Specifically in the main problem of following section, we consider the buffer constraint of $\mathit{Q}[n-1]+\mathit{C}_{\mathcal{SR}}[n]\delta_{t}-\mathit{C}_{\mathcal{RD}}[n]\delta_{t} \leq L_{Q}, \forall n$, which prevents the overflow issue.
Note that $L_{\mathrm{Q}}$ denotes the size of buffer. }
The normalized remaining bits in the buffer of relay evolves according to

\begin{equation}
\mathit{Q}[n] = \mathit{Q}[n-1]+\mathit{C}_{\mathcal{SR}}[n]\delta_{t}-\mathit{C}_{\mathcal{RD}}[n]\delta_{t}, \ n=1,2, \cdots, N, \label{Q[n]}
\end{equation}
where $\mathit{Q}[0]$ and $\mathit{C}_{\mathcal{RD}}[1]$ are equal to zero.
Note that $\mathit{C}_{\mathcal{RD}}[n]$ denotes the data rate received by the destination (i.e., the user terminal) in time slot $n$.
Accordingly, it evolves to 

\begin{equation}
\mathit{Q}[n] = \displaystyle\sum_{i=1}^{n} \mathit{C}_{\mathcal{SR}}[i]\delta_t - \displaystyle\sum_{i=2}^{n} \mathit{C}_{\mathcal{RD}}[i]\delta_{t} , \ n=2,3, \cdots, N. \label{Q[n]_0}    
\end{equation}
Note that $\mathit{Q}[1]= \mathit{C}_{\mathcal{SR}}[1] \delta_t$.

\subsubsection{Destination receives}
The maximum amount of transmittable bits at $\mathcal{R}$ is limited by the remaining bits in the buffer or the instantaneous capacity for $\mathcal{R-D}$ link.
Thus, the transmission rate of $\mathcal{R-D}$ link in time slot $n$ in bps is given by

\begin{eqnarray}
\mathit{C}_{\mathcal{RD}}[n] = \mathrm{min}\lbrace R_{\mathrm{RF}}[n],\mathit{Q}[n-1]/\delta_t+\mathit{C}_{\mathcal{SR}}[n] \rbrace, & & \nonumber \label{C_RD} \\
n = 2, \cdots, N. & &
\end{eqnarray}

The conventional relaying introduces a delay of one time slot since the relay needs to wait until the entire data is received and decoded before forwarding the data to the destination, especially in DF protocol.
The relay in this system receives data from the source in the current time slots and sends this cumulative information to the destination in the next time slots.
We thus consider $\mathit{C}_{\mathcal{RD}}[n]$ over time slots $n = 2, \cdots, N$ (i.e., $\mathit{C}_{\mathcal{RD}}[1]=0$).

Subsequently, the average throughput in the mixed FSO/RF communication under the buffer constraint is given by 

\begin{eqnarray}
\Phi &=& \frac{1}{N-1}\sum_{n=2}^{N}\mathit{C}_{\mathcal{RD}}[n]. \label{AverageThroughput} 
\end{eqnarray}

The delay in the system is defined by the duration between the time when the bits leave the source node and the time when it arrives at the destination. 
We note that the average delay is proportional to the average queue length for a given arrival rate from Little’s Theorem \cite{System_10, System_6}. 
As a result, the average queue length can bridge the average delay. 
Thus, we address the average delay at relay $\mathcal{R}$ by the average queue length as following

\begin{eqnarray}
& & L = \frac{ \mathbb{E}\lbrace \mathit{Q[n]} \rbrace }{\lambda}. 
\label{average queue length} 
\end{eqnarray}
Note that the average arrival rate of bits per slot into the queue of the buffer denoted by $\lambda$ is defined as $\mathbb{E}\lbrace \mathit{\mathit{C}_{\mathcal{SR}}[n]}\rbrace$.
Since it takes one time slot to transmit data from the source to a relay node, the average delay in the system is given by
$\bar{L} = L - \delta_t$.
To account for the delay-considered application, the average delay is fundamental; therefore, we specifically consider the delay constraint of $L \leq L^{\mathrm{req}}$ in the main problem of following section.

In practice, there are usually some constraints on the delay and on the buffer size. In the following section, these constraints are investigated in the proposed mixed FSO/RF mobile relaying system. 
\tred{
For the three-node network considered, we assume that the source always has information to transmit, and hence the end-to-end delay is mainly caused by a buffer in the relay.}
Based on the given system model, our objective of the following section is the maximization of average throughput $\Phi$ by efficient trajectory design, taking account of a limited buffer under a delay constraint.

\tred{
\section{Throughput Maximization for Buffer-aided Mobile Relaying under Delay-Consideration}
}\label{Body}

In order to find the efficient trajectory of UAV-aided relay maximizing the throughput of mixed FSO/RF under the consideration of limited buffer size and delay-limited, the following two problems are formulated.
Motivated by employing mobile relay to provide both delay-limited and delay-tolerant services in future wireless networks as in \cite{Intro_9}, we find the optimal delay-considered policies which study not only buffer requirement but also delay requirement.
We will investigate the delay-limited transmission case. 
As a special case of the delay-limited transmission, the delay-tolerant transmission case \cite{ICC} is also studied.

\subsection{Problem Formulation} \label{section:Delay-Limited Transmission} 
The throughput maximization problem for this transmission scheme should consider a delay in a queue; hence, we use the average delay $L$ in \eqref{average queue length} for the delay requirement.

Since the average throughput $\Phi$ can be expressed equivalently with a sum of throughput received by the destination, we set the objective function with  $\sum_{n=2}^{N}\mathit{C}_{\mathcal{RD}}[n]$.
Note that we adopt the following notations to better understand the continuous variables in the optimization problems: the position of UAV $\mathcal{Q} = \{\mathbf{q}_{\mathcal{R}}[n], \ \forall n\}$, the velocity of UAV $\mathcal{V} = \{\mathbf{v}_{\mathcal{R}}[n], \ \forall n\}$, and the acceleration of UAV $\mathcal{A} = \{\mathbf{a}_{\mathcal{R}}[n], \ \forall n\}$.
Thus, we formulate the throughput maximization for delay-limited transmission as the following (P1):

\begin{eqnarray}
(\mathrm{P1}) \!\!&\!\!	\displaystyle \max_{ \scriptsize \begin{array}{c} \scriptsize \mathcal{Q}, \mathcal{V}, \mathcal{A}  \end{array} } \!\!&\!\! 
	\sum_{n=2}^{N} \mathit{C}_{\mathcal{RD}}[n] \nonumber \label{P1}\\  			
\!\!\!\!&\!\!\!\!	\textrm{s.t} \!\!\!\!&\!\!\!\! \mathbf{v}_{\mathcal{R}}[n+1]=\mathbf{v}_{\mathcal{R}}[n]+\mathbf{a}_{\mathcal{R}}[n]\delta_t, \nonumber \\
\!\!\!\!&\!\!\!\! \!\!\!\!&\!\!\!\!	\mathbf{q}_{\mathcal{R}}[n+1]=\mathbf{q}_{\mathcal{R}}[n]+\mathbf{v}_{\mathcal{R}}[n]\delta_t + \frac{1}{2}\mathbf{a}_{\mathcal{R}}[n]\delta_t^2, \nonumber \\
\!\!\!\!&\!\!\!\! \!\!\!\!&\!\!\!\! n=0,1, \cdots, N, \label{C_q&v&a_P0} \\
\!\!\!\!&\!\!\!\! \!\!\!\!&\!\!\!\!	\|\mathbf{a}_{\mathcal{R}}[n]\| \leq A_{\max}, \ n=0,1, \cdots, N, \label{C_Amax_P0} \\
\!\!\!\!&\!\!\!\! \!\!\!\!&\!\!\!\!	\|\mathbf{v}_{\mathcal{R}}[n]\| \leq V_{\max}, \ n=1,2, \cdots, N, \label{C_Vmax_P0} \\
\!\!\!\!&\!\!\!\! \!\!\!\!&\!\!\!\! 0 \leq \mathit{C}_{\mathcal{SR}}[n], \ n=1,2,\cdots, N, \label{C_Csr_P1} \\
\!\!\!\!&\!\!\!\! \!\!\!\!&\!\!\!\! 0 \leq \mathit{Q}[n] \leq L_{\mathrm{Q}}, \ n=1,2,\cdots, N, \label{C_Q0_P1} \\
\!\!\!\!&\!\!\!\! \!\!\!\!&\!\!\!\! L \leq L^{\mathrm{req}}. \label{C_L_P1} 
\end{eqnarray}

The constraint in \eqref{C_q&v&a_P0} describes the discrete time model of UAV's position and velocity related to the position $\mathbf{q}_{\mathcal{R}}[n]$, the velocity $\mathbf{v}_{\mathcal{R}}[n]$, as well as the acceleration $\mathbf{a}_{\mathcal{R}}[n]$.
Moreover, considering practical constraints of UAV's flight, UAV is constrained with the maximum acceleration in \eqref{C_Amax_P0}, and the maximum velocity in \eqref{C_Vmax_P0}.
In addition, we establish \eqref{C_Q0_P1} to hold the buffer constraint, which limits the queue length to $L_{\mathrm{Q}}$.
Note that \eqref{C_Q0_P1} satisfies the information-causality by $\mathit{Q}[n] \geq 0$\footnote{
$\mathcal{R}$ can only forward the information which has already been received from $\mathcal{S}$, at each slot $n$.
}.
Furthermore, delay-considered transmission yields the average delay constraint of \eqref{C_L_P1}, which includes the requirement value of the average delay $L^{\mathrm{req}}$. 
Depending on a certain delay-requirement, the delay-time limit can be flexibly managed by adjusting $L^{\mathrm{req}}$.

\tred{
We note that, in some applications (e.g., sensing and border surveillance), UAV should be constrained with pre-determined initial and final positions/velocities. 
If necessary, the initial and final position/velocity constraints can be further considered in this framework (P1) as follows
}
\begin{eqnarray}
&&\mathbf{q}_{\mathcal{R}}[0]=\mathbf{q}_{\mathrm{I}}, \  \mathbf{q}_{\mathcal{R}}[N+1]=\mathbf{q}_{\mathrm{F}}, \label{qI} \\ &&\mathbf{v}_{\mathcal{R}}[0]=\mathbf{v}_{\mathrm{I}}, \  \mathbf{v}_{\mathcal{R}}[N+1]=\mathbf{v}_{\mathrm{F}}, \label{qF}
\end{eqnarray}
\tred{
In \eqref{qI} and \eqref{qF}, we denote $\mathbf{q}_{\mathrm{I}}$, $\mathbf{q}_{\mathrm{F}}$, $\mathbf{v}_{\mathrm{I}}$, and $\mathbf{v}_{\mathrm{F}}$ as the initial/final positions and velocities, respectively.
In this system, we do not consider these constraints to solely focus on the mobile relaying system, especially for mixed FSO/RF-enabled backhaul networks.
}

\subsection{Proposed Algorithm}
Despite of the convex constraints \eqref{C_q&v&a_P0}-\eqref{C_Vmax_P0}, the non-concave objective function $\mathit{C}_{\mathcal{RD}}[n]$ and non-convex constraints \eqref{C_Q0_P1}-\eqref{C_L_P1} cause (P1) to be a non-convex optimization problem, which therefore is quite challenging to solve  with the standard convex optimization method.
To address such a challenge, we first use the first-order Taylor approximation to $R_{\mathrm{RF}}[n]$ with any given local value $\mathbf{q}^{k}_{\mathcal{R}}[n]$ at the iteration $k$.
Then, we lower-bound the throughput of RF as

\begin{equation}
\begin{split}
R^{k}_{\mathrm{RF}}[n] =& \ B_{\mathrm{RF}} \cdot \big( \mathcal{A}^{k} - \mathcal{B}^{k} ( \|\mathbf{q}_{\mathcal{D}} - \mathbf{q}_{\mathcal{R}}[n]\|^2 \\
&- \|\mathbf{q}_{\mathcal{D}} - \mathbf{q}^{k}_{\mathcal{R}}[n]\|^2 )\big), \  n = 2,3,\cdots, N. 
\end{split}
\end{equation}
Note that, as in \cite{Body_1}, $\mathcal{A}^{k}$ and $\mathcal{B}^{k}$ are expressed as

\begin{eqnarray}
\!\!\!\!\!\!&\!\!\!\!\!\!&\!\!\!\!\!\! \mathcal{A}^{k}=\frac{1}{\log 2} \cdot \mathrm{log} \left(1 + \frac{\gamma_{0}}{(\|\mathbf{q}_{\mathcal{D}} - \mathbf{q}^{k}_{\mathcal{R}}[n] \|^2)^{\alpha}} \right), \\
\!\!\!\!\!\!&\!\!\!\!\!\!&\!\!\!\!\!\! \mathcal{B}^{k}=\frac{\gamma_{0} \alpha}{\log 2 \cdot (\gamma_{0}+(\|\mathbf{q}_{\mathcal{D}} - \mathbf{q}^{k}_{\mathcal{R}}[n] \|^2)^{\alpha})(\|\mathbf{q}_{\mathcal{D}} - \mathbf{q}^{k}_{\mathcal{R}}[n]\|^{2})},  \nonumber \\
&\!\!\!\!\!\!&\!\!\!\! n = 2,3,\cdots, N. 
\end{eqnarray}

Secondly, the lower bounded throughput of FSO is given by high-SNR approximation\footnote{Since $k_1 \cdot e^{ -k_2 \cdot \|\mathbf{q}_{\mathcal{R}}[n] - \mathbf{q}_{\mathcal{S}} \| }\gg1$ even under the worst atmospheric conditions (e.g., heavy-fog condition), the high-SNR approximation can be applied to the achievable rate of FSO link.
} as 

\begin{equation}
\begin{split}
R^{k}_{\mathrm{FSO}}[n] = \frac{B_{\mathrm{FSO}}}{2\log{2}} \left(\log(k_1) -k_2 \cdot \| \mathbf{q}_{\mathcal{R}}[n] - \mathbf{q}_{\mathcal{S}} \| \right), &  \\ 
n=1,2,\cdots, N. &
\end{split}
\end{equation}
Note that $R^{k}_{\mathrm{RF}}[n]$ and $R^{k}_{\mathrm{FSO}}[n]$ are concave functions with respect to $\mathbf{q}_{\mathcal{R}}[n]$.

Now, we introduce slack variables $\mathcal{T_S} = \{t_{\mathcal{S}}[n]= \mathit{C}_{\mathcal{SR}}[n], \ n = 1, \cdots, N \}$, $\mathcal{T_D} = \{t_{\mathcal{D}}[n]= \mathit{C}_{\mathcal{RD}}[n], \ n = 2, \cdots, N \}$ for the non-concave objective function and replace the non-convex constraints \eqref{C_Q0_P1}-\eqref{C_L_P1} to convex constraints with $\lbrace t_{\mathcal{S}}[n] \rbrace_{n=1}^{N}$ and $\lbrace t_{\mathcal{D}}[n] \rbrace_{n=2}^{N}$.
Hence, we can reformulate (P1) into the following optimization problem for any given local value $\lbrace \mathbf{q}^{k}_{\mathcal{R}}[n] \rbrace_{n=1}^{N}$ at the $k$-th iteration as following

\begin{eqnarray}
\!\!\!\!\!\! (\mathrm{P1}^\star) \!\!\!\!&\!\! \displaystyle\max_{ \scriptsize \begin{array}{c} 
\mathcal{T_S}, \mathcal{T_D}, \\ 
\mathcal{Q}, \mathcal{V}, \mathcal{A}	 
\end{array} }   \!\!\!\!\!&\! \ \sum_{n=2}^{N} t_{\mathcal{D}}[n]  \nonumber \label{P2}\\   \!&\! \	\textrm{s.t}  \!\!\!\!\!&\! \ \eqref{C_q&v&a_P0}-\eqref{C_Vmax_P0},  \nonumber \\
\!&\!\!&\! 0 \leq t_{\mathcal{S}}[n] \leq R^{k}_{\mathrm{FSO}}[n], n=1,2,\cdots, N, \label{C_tS_P2} \\
\!&\!\!&\! 0 \leq t_{\mathcal{D}}[n] \leq R^{k}_{\mathrm{RF}}[n], n=2,3,\cdots, N, \label{C_tD_1_P2} \\
\!&\!\!&\! t_{\mathcal{D}}[n] \leq \mathit{Q}^{'}[n-1]/\delta_t +t_{\mathcal{S}}[n], \nonumber \\
\!&\!\!&\! n=2,3,\cdots, N, \label{C_tD_2_P2} \\
\!&\!\!&\! \mathit{Q}^{'}[n]\geq 0, n=1,2,\cdots, N, \label{C_Q0_1_P2} \\
\!&\!\!&\! \mathit{Q}^{'}[n] \leq L_{Q}, n=1,2,\cdots, N, \label{C_Q0_2_P2} \\
\!&\!\!&\! \sum_{n=1}^{N} \mathit{Q}^{'}[n] \leq L^{\mathrm{req}} \sum_{n=1}^{N} t_{\mathcal{S}}[n], \label{C_L_P2} 
\end{eqnarray}
where we note that, plugging \eqref{C_Q0_P1} into \eqref{Q[n]}, we can recursively rewrite the queue state as  

\begin{equation}
\mathit{Q}^{'}[n] = \displaystyle\sum_{i=1}^{n} t_{\mathcal{S}}[i]\delta_t - \displaystyle\sum_{i=2}^{n} t_{\mathcal{D}}[i]\delta_t , \ n=2,3, \cdots, N, \label{Q[n]_1}
\end{equation}
in which $\mathit{Q}^{'}[1]= t_{\mathcal{S}}[1]$.
We also point out that the average delay is rewritten as $L^{'} = \frac{ \sum_{n=1}^{N} \mathit{Q}^{'}[n] }{\sum_{n=1}^{N} t_{\mathcal{S}}[n]}$ yielding the constraint in \eqref{C_L_P2}.


$(\mathrm{P1}^\star)$ is the type of convex quadratically constrained program (QCP).
The convex QCP can be tackled within a polynomial complexity, by interior-point methods with a standard convex optimization solver (e.g., CVX). 
Therefore, we suboptimally solve (P1) via the successive convex approximation (SCA) method to (P1$^\star$) by iteratively updating the local values $\lbrace \mathbf{q}^{k}_{\mathcal{R}}[n] \rbrace_{n=1}^{N}$ \cite{Intro_8}. 
It has been proved that the SCA method converges to at least a locally optimal point \cite{Body_3}.
We further show the convergence through the numerical results in Section \ref{Numerical Result}.

In closing this subsection, we summarize the proposed successive optimization steps for the delay-limited transmission (P1) in \textbf{Algorithm \ref{algorithm 1}}.
\begin{algorithm}[h] 
  \small
  \caption{Proposed Algorithm for Throughput Maximization with a Limited Buffer} \label{algorithm 1}
  \Input{$L_{Q}$, $L^{\mathrm{req}}$, $B_{\mathrm{RF}}$, $B_{\mathrm{FSO}}$, $\gamma_{0}$, $\gamma_{\mathrm{FSO}}$, $V$, and a set of parameters related to UAV's flight}
  \Output{Optimized values of $\lbrace \mathbf{q}_{\mathcal{R}}[n] \rbrace_{n=1}^{N}$, $\lbrace \mathbf{v}_{\mathcal{R}}[n] \rbrace_{n=1}^{N}$, $\lbrace \mathbf{a}_{\mathcal{R}}[n] \rbrace_{n=0}^{N}$, $\lbrace t_{\mathcal{S}}[n] \rbrace_{n=1}^{N}$, and $\lbrace t_{\mathcal{D}}[n] \rbrace_{n=2}^{N}$
  }
  \vspace{0.05in}
  Initialize the UAV's position vector $\lbrace \mathbf{q}^{0}_{\mathcal{R}}[n] \rbrace_{n=1}^{N}$, and set the iteration number $k=0$ \;
  \While{the partial increase for the objective value of $(\mathrm{P1}^\star)$ is above a tolerance $\varepsilon$,}{
    Find the optimal solution $\lbrace \mathbf{q}^{*}_{\mathcal{R}}[n] \rbrace_{n=1}^{N}$  to (P1$^\star$) for the local values $\lbrace \mathbf{q}^{k}_{\mathcal{R}}[n] \rbrace_{n=1}^{N}$ at the iteration $k$\; 
  	Update $k=k+1$\;    
    Update the optimal solution as $\mathbf{q}^{k}_{\mathcal{R}}[n]=\mathbf{q}^{*}_{\mathcal{R}}[n]$, $n=1,2,\cdots,N$\;
  }
\end{algorithm}

\subsection{Case Study: Delay-Tolerant Transmission} \label{section:Delay-Tolerant Transmission} 
Previously, we have dealt with the delay-limited transmission scheme, and have studied throughput optimization on UAV-assisted mobile relaying under a limited buffer constraint.
Clearly, most applications focus on the throughput and delay, which have a trade-off between them (see, e.g., \cite{System_6, System_7}).
In contrast, some applications, such as periodic sensing, do not concern the delay as sensitively as the delay-limited application does.
Hence, we address the delay-tolerant transmission scheme as a case study of (P1).

For delay-tolerant transmission scheme, the relay is allowed to store the received data in its buffer and forward it to the destination without any limit of delay.
Accordingly, we can assume that the delay requirement is very loose, e.g., $L^{\mathrm{req}}$ is very large in the delay-tolerant transmission. 
Then, we can omit \eqref{C_L_P1} since it does not affect the throughput optimization. 
In particular, we deal with the delay-tolerant transmission scheme by simply changing the constraint in $(\mathrm{P1}^\star)$ as follows:

\begin{eqnarray}
(\mathrm{P2}^\star) \!&\! \displaystyle\max_{ \scriptsize \begin{array}{c} 
\mathcal{T_S}, \mathcal{T_D}, \\ 
\mathcal{Q}, \mathcal{V}, \mathcal{A}	 
\end{array} }   \!&\! \ \sum_{n=2}^{N} t_{\mathcal{D}}[n]  \nonumber \label{P2_1}\\   \!&\! \	\textrm{s.t}   \!&\! \ \eqref{C_q&v&a_P0}-\eqref{C_Vmax_P0}, \eqref{C_tS_P2}-\eqref{C_Q0_2_P2},  \nonumber \\
\!&\!\!&\! L^{'} \leq \infty. \label{C_L_P3} 
\end{eqnarray}

Like the problem $(\mathrm{P1}^\star)$, the problem $(\mathrm{P2}^\star)$ in \eqref{C_L_P3} is also QCP.
Then, we suboptimally solve (P2$^\star$) via the successive convex optimization by iteratively updating the local points $\lbrace \mathbf{q}^{k}_{\mathcal{R}}[n] \rbrace_{n=1}^{N}$, which ensure to converge. 
Note that we can also apply \textbf{Algorithm \ref{algorithm 1}} to the throughput maximization of delay-tolerant transmission by replacing $(\mathrm{P1}^\star)$ to $(\mathrm{P2}^\star)$ and \eqref{C_L_P2} to \eqref{C_L_P3}.

\subsection{Complexity Analysis}
We here provide the complexity of our algorithm. 
In order to determine the complexity of \textbf{Algorithm \ref{algorithm 1}}, we need to decide the complexity of subproblem (P1$^\star$) described in Section \ref{section:Delay-Limited Transmission}.

According to the complexity analysis in \cite{Body2_4, Body2_5}, to solve the convex optimization problem especially with the interior-point methods, (ignoring any structure in the problem, such as sparsity) each step requires on the order of 

\begin{equation}
    \mathrm{max}\{ \zeta^3, \zeta^2\xi, F \}
\end{equation}
operations. 
Note that $\zeta$ and $\xi$ denote the number of variables and constraints, respectively, and $F$ denotes the cost of evaluating the first and second derivatives of the objective and constraint functions.

For (P1$^\star$), it can be easily found that $\zeta = 8N+1$.
Also, we can compute the number of constraints $\xi$, according to Table \ref{table:b} at the top of next page, as $\xi = 11N + 4$.
Therefore, comparing the order of $\zeta^3$, $\zeta^2\xi$, and $F$, it can be found that $\zeta^2\xi$ is greater than $\zeta^3$ and $F$.
Note that $F$ follows $O(N)$ in the problems, whereas, $\zeta^3$ and $\zeta^2\xi$ follow $O(N^3)$.

As a result, considering the interior point method, the computational complexity for Algorithm \ref{algorithm 1} is derived as 

\begin{equation}
\sum^{\vartheta}_{m=1}{704N^3 + 432N^2 + 75N + 4}.  
\end{equation}
Note that $\vartheta$ denotes the number of iterations for steps 3 - 5 in Algorithm \ref{algorithm 1}.  
Then, the complexity of Algorithm \ref{algorithm 1} can be reduced to the order of $O(\vartheta N^3)$.



\begin{table}[t]\footnotesize
	\centering
	\caption{\tred{Number of constraints in (P1$^\star$).}}
	\resizebox{0.65\columnwidth}{!}	{\begin{tabular} {c |c| c| c| c| c| c| c| c| c}
			\toprule[1pt]
			\textbf{Equation} & \eqref{C_q&v&a_P0} & \eqref{C_Amax_P0} & \eqref{C_Vmax_P0} & \eqref{C_tS_P2} & \eqref{C_tD_1_P2} &  \eqref{C_tD_2_P2} & \eqref{C_Q0_1_P2}  & \eqref{C_Q0_2_P2} & \eqref{C_L_P2}   \\
			\midrule
			\textbf{The number of } & \multirow{2}{*}{$4(N+1)$} & \multirow{2}{*}{$N+1$} & \multirow{2}{*}{$N$} & \multirow{2}{*}{$N$} & \multirow{2}{*}{$N-1$} & \multirow{2}{*}{$N-1$} & \multirow{2}{*}{$N$} & \multirow{2}{*}{$N$} & \multirow{2}{*}{$1$} \\
			\textbf{constraints} & & & & & & & & &\\ 
			\bottomrule[1pt]
	\end{tabular}}
	\label{table:b}
\end{table}


\section{Numerical Results} \label{Numerical Result}

In this section, some selected numerical results are provided to validate the proposed algorithm and evaluate the mixed FSO/RF UAV-enabled mobile relaying system with a buffer. 
In particular, the simulation results for delay-limited transmission are presented, taking account of the average delay requirement.
Then, to consider a general case of delay-tolerant application, the simulation results for delay-tolerant transmission are further presented. 
Lastly, we compare and analyze the proposed scheme with conventional schemes.

\tred{
A system is assumed with the fixed altitude of UAV $H=100$ [m] which is possible lowest altitude by regulation}, the location of backhaul terminal $\mathbf{q}_{\mathcal{S}}=[0,0,0]^{T}$, the location of user terminal $\mathbf{q}_{\mathcal{D}}=[L,0,0]^{T}$ where $L=2000$ [m].
For the UAV-aided mobile relaying system, we consider that the maximum velocity $V_{\mathrm{max}}=50$ [m/s], and the maximum acceleration $A_{\mathrm{max}}=5$ [m/$\mathrm{s}^2$]. 
Unless stated otherwise, we consider the period $T=200$ [s], 
the light fog condition of visibility $V=0.8$ [km], the bandwidth for FSO $B_{\mathrm{FSO}}=10^8$ [Hz], the bandwidth for RF $B_{\mathrm{RF}}=10^8$ [Hz], the ASNR $\gamma_{\mathrm{FSO}} = 5$ [dB] $(\alpha_{0}=\frac{1}{10})$, and the reference received SNR $\gamma = 6$ [dB]. 
Note that $\gamma$ represents the reference received SNR for $\mathcal{R-D}$ link when the UAV is located above the user terminal. 
The parameters for the probabilistic LoS channel model in $P_{\mathrm{LoS}}[n]$ are set as $C = 10$, $D = 0.6$, $\kappa = 0.2$, and $\alpha = 2.2$. 
Moreover, the regularized homogeneous LoS probability $\bar{P}_{\mathrm{LoS}}$ in \eqref{Rate_RF_4} is set as the value corresponding to the elevation angle of $90^{\circ}$.
The simulation results of this paper are obtained through CVX.

\subsection{Simulation Results for Delay-Limited Transmission}
\begin{figure*}
    \vspace{-1em}
    \centering
    \includegraphics[width=0.75\textwidth]{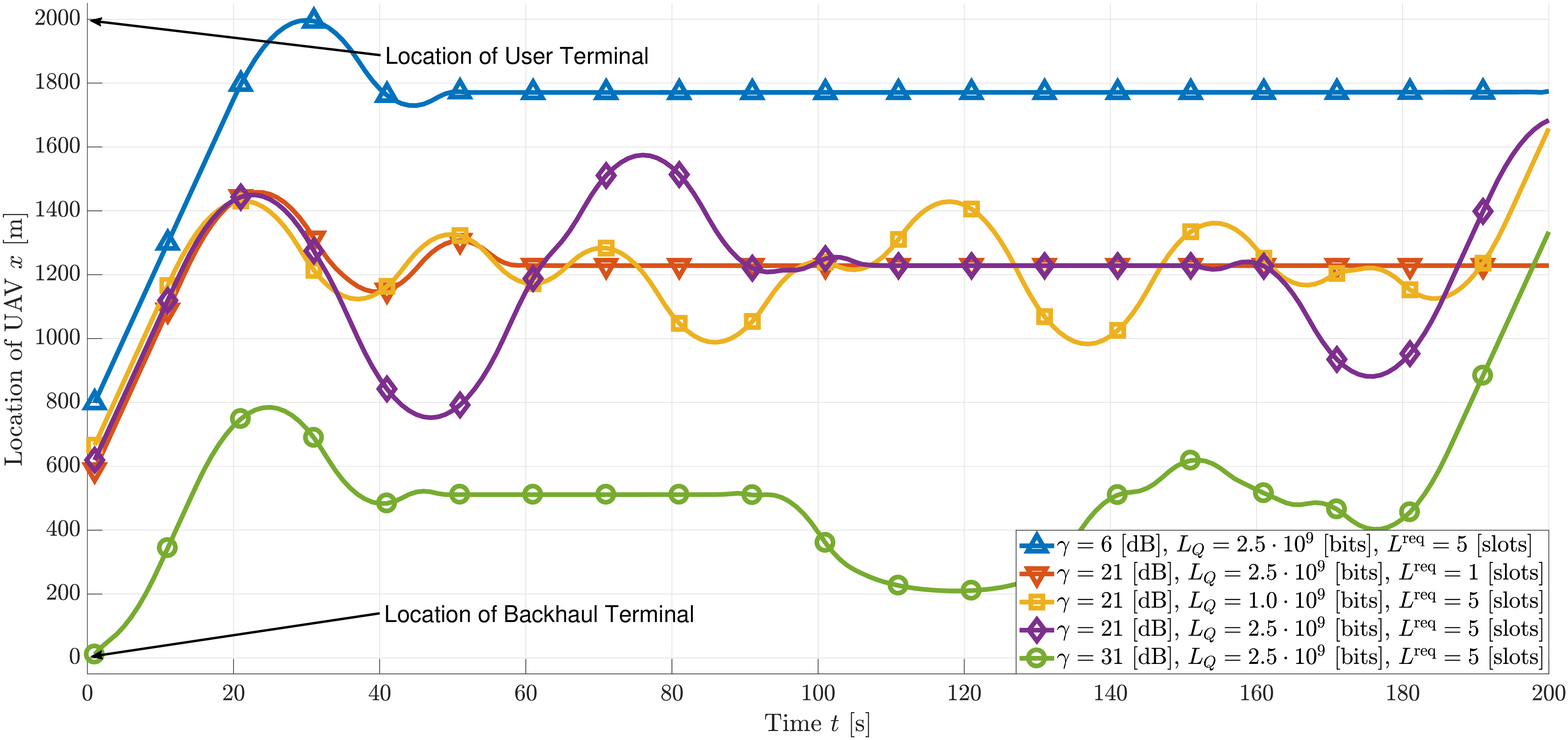}
    \caption{Trajectory comparison with optimized $x$-coordinate position over time $t$ for different reference received SNR, buffer size, and delay requirement conditions.}
    \label{fig:Xplot_gamma&LQ&Lreq}
    \vspace{-2.0em}
\end{figure*}

\begin{figure}
    \centering
    \includegraphics[width=0.48\textwidth]{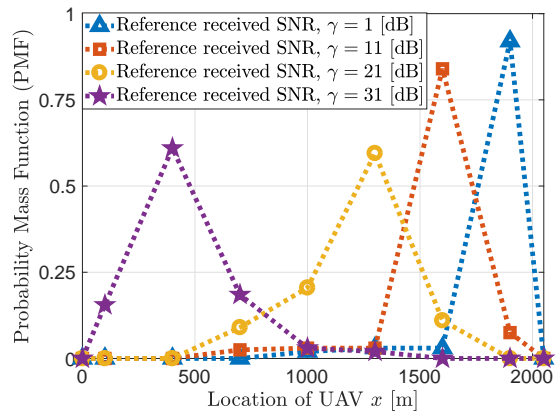}
    \caption{PMF based on $200$ [m] interval for throughput maximized trajectories over reference received SNR $\gamma$.}
    \label{fig:Histogram_gamma}
\end{figure}

Considering the average delay requirement $L^{\mathrm{req}}$, we present the simulation results of the throughput maximization problem for delay-limited transmission (P1). 
In Fig. \ref{fig:Xplot_gamma&LQ&Lreq}, the optimized trajectories in $x$-coordinate\footnote{
The backhaul terminal and user terminal are positioned horizontally on the $x$-axis. 
Intuitively, we can see that flying between backhaul and user terminals only on the $x$-axis is optimal, i.e., any UAV movement on the y-axis is not beneficial. 
Thus, we provide only the result of the $x$-axis, not the trajectory of $xy$-axis or $xyz$-axis.
} for the mobile relay are drawn over the different reference received SNR $\gamma$, buffer size $L_{Q}$, and delay requirement $L^{\mathrm{req}}$.
It can be found that the UAV circulates (or stays) between $\mathcal{S}$ and $\mathcal{D}$, to deliver the data by efficiently balancing storing and forwarding data in the buffer when the FSO and RF links are comparable. 
For example, under ``$\gamma=21$ [dB], $L_{Q}=2.5 \cdot 10^{9}$ [bits], $L^{\mathrm{req}}=5$ [slots]" in Fig. \ref{fig:Xplot_gamma&LQ&Lreq}, the UAV circulates about seven times around $x = 1219$ [m].
On the other hands, by limiting to less buffer size or tightening the delay requirement, optimal trajectory is drawn in the different pattern as seen from the comparison of ``$\gamma=21$ [dB], $L_{Q}=2.5 \cdot 10^{9}$ [bits], $L^{\mathrm{req}}=5$ [slots]" with ``$\gamma=21$ [dB], $L_{Q}=1.0 \cdot 10^{9}$ [bits], $L^{\mathrm{req}}=5$ [slots]" and ``$\gamma=21$ [dB], $L_{Q}=2.5 \cdot 10^{9}$ [bits], $L^{\mathrm{req}}=1$ [slots]", respectively.
As the smaller buffer size or tighter delay requirement is given, UAV tends to stay at a certain position, since data in the buffer needs to be de-queued faster.

\begin{figure}[]
    \centering
    \includegraphics[width=0.48\textwidth]{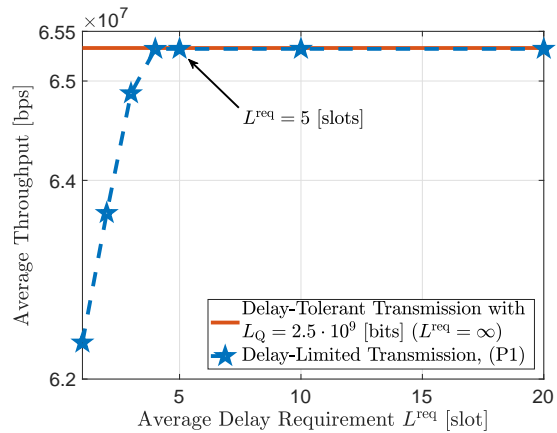}
    \caption{Average throughput over average delay limit $L^{\mathrm{req}}$ [slot].}
    \label{fig:Throughput_L}
\end{figure}

Furthermore, Figs. \ref{fig:Xplot_gamma&LQ&Lreq} and \ref{fig:Histogram_gamma} show that, as $\gamma$ increases, the optimal UAV trajectory moves from the users terminal toward the backhaul terminal with the given buffer size $L_{\mathrm{Q}}=2.5 \cdot 10^{9}$ [bits] (which is the optimal buffer size [bits]\footnote{The optimal buffer size and the optimal delay requirement depend on the set of parameters related to communication conditions of each link. \label{footnote1}
} verified in Fig. \ref{fig:Throughput_BUF}) and delay requirement $L^{\mathrm{req}}=5$ [slots] (which is the optimal delay requirement [slots]\footref{footnote1} verified in Fig. \ref{fig:Throughput_L}).
For instance, in Fig. \ref{fig:Histogram_gamma} which presents the probability mass function (PMF) based on $300$ [m] interval for the different $\gamma$, the UAV mainly stays near the user terminal ($1450$ [m] $\sim 1750$ [m]) about 84\% of the total flight time with $\gamma=11$ [dB], while staying near the backhaul terminal ($250$ [m] $\sim 550$ [m]) about 61\% of the total flight time with $\gamma=31$ [dB].
One can observe that how long UAV stays in a certain interval between the backhaul terminal and the user terminal from the result of PMF.

\begin{figure}[]
    \centering
    \includegraphics[width=0.48\textwidth]{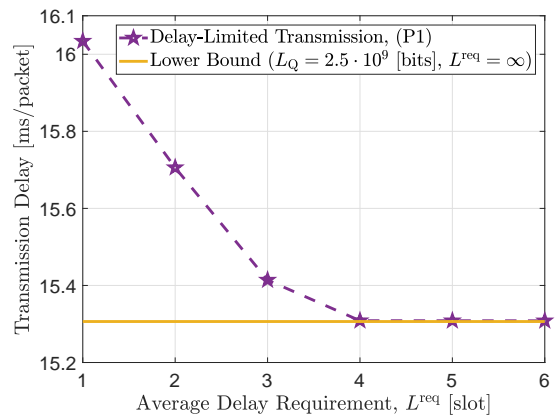}
    \caption{\tred{Transmission delay over average delay limit $L^{\mathrm{req}}$ [slot].}}
    \label{fig:Delay_L}
\end{figure}

\tred{
Fig. \ref{fig:Throughput_L} presents the optimal average throughput of delay-limited transmission (P1) over the average delay limit.
We can see that as $L^{\mathrm{req}}$ increases, the average throughput increases.
Since larger $L^{\mathrm{req}}$ given, the path for UAV can be designed more flexibly to maximize the average throughput of mobile relays (i.e., less constrained by \eqref{C_L_P2}).  
It is also observed that the average throughput results are close enough to the upper bound of (P1) given an average delay requirement larger than $L^{\mathrm{req}}=5$ [slots].
Note that the result of ``Delay-Tolerant Transmission with $L_{\mathrm{Q}}=2.5 \cdot 10^{9}$ [bits] ($L^{\mathrm{req}}=\infty$)" in Fig. \ref{fig:Throughput_L}, which considers no delay limit, can be used as the upper bound of (P1).
Thus, we can confirm that $L^{\mathrm{req}}=5$ [slots] is the optimal average delay limit size in terms of throughput.
It is worth noting that Fig. \ref{fig:Throughput_L} also shows the throughput-delay tradeoff which indicates that throughput and delay are two conflict metrics, and thus, it is necessary to balance them effectively.}
As a result, we obtain the optimal buffer size $L_{\mathrm{Q}}=2.5 \cdot 10^{9}$ [bits] (which verified for the optimal buffer size [bits] in Fig. \ref{fig:Throughput_BUF}) and the optimal delay requirement $L^{\mathrm{req}}=5$ [bits] for the throughput maximization.
We note that the obtained optimal buffer size and optimal delay requirement depend on the given system and channel parameters (e.g., weather condition, conditions of FSO link $\gamma_{\mathrm{FSO}}$ and $B_{\mathrm{FSO}}$, and conditions of RF link $\gamma$ and $B_{\mathrm{RF}}$).
\tred{
In addition to the throughput result of (P1) in Fig. \ref{fig:Throughput_L},  we also present the delay result of (P1) in Fig. \ref{fig:Delay_L}.
As in \cite{Num_7}, we consider that transmission delay is the packet delay between source $\mathcal{S}$ and destination $\mathcal{D}$ in which the packet size is $10^{6}$ [bits], and the queuing delay is the time when a job waits in a queue in relay $\mathcal{R}$ until it can be executed.
It is observed that the transmission delay decreases and then converges as larger delay requirements are given.}

\begin{figure}[]
    \centering
    \includegraphics[width=0.48\textwidth]{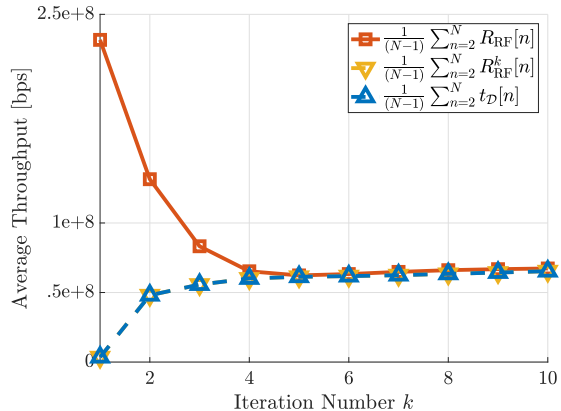}
    \caption{Convergence plot of Algorithm \ref{algorithm 1}.}
    \label{fig:Convergence_L}
\end{figure}

Fig. \ref{fig:Convergence_L} shows the convergence of Algorithm \ref{algorithm 1}.
Especially, this figure shows how much the slack variable $t_{\mathcal{D}}$, which is introduced for non-convexity of (P1), approaches to the throughput for RF link $R_{\mathrm{RF}}[n]$ and the lower-bounded throughput for RF link $R_{\mathrm{RF}}^{k}[n]$ according to the number of iterations in Algorithm \ref{algorithm 1}.
Note that the result of Fig. \ref{fig:Convergence_L} yields from (P1) with $L_{\mathrm{Q}}=2.5 \cdot 10^{9}$ [bits] and $L^{\mathrm{req}}=5$ [slots].
As shown in this figure, $t_D$ and $R_{\mathrm{RF}}^{k}[n]$ are close enough to $R_{\mathrm{RF}}[n]$ after five iterations, and the local optimal solution of $(\mathrm{P1}^\star)$ in Algorithm \ref{algorithm 1} converges by the iteration.

\subsection{Simulation Results for Delay-Tolerant Transmission}

Accounting for the case study in Section \ref{section:Delay-Tolerant Transmission}, we present the numerical result for the buffer size constraint.

\begin{figure}[]
    \centering
    \includegraphics[width=0.48\textwidth]{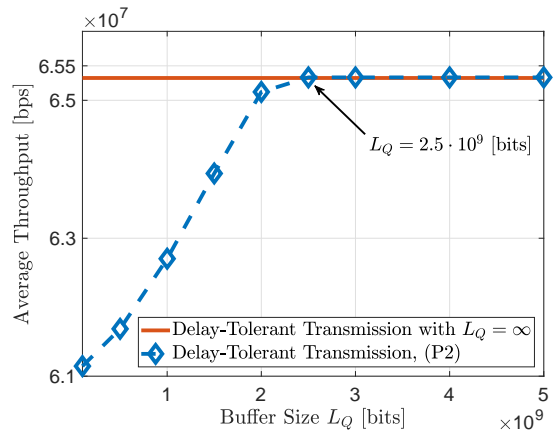}
    \caption{Average throughput [bps] over buffer size $L_{\mathrm{Q}}$ [bits].}
    \label{fig:Throughput_BUF}
\end{figure}

\begin{figure}[]
    \centering
    \includegraphics[width=0.48\textwidth]{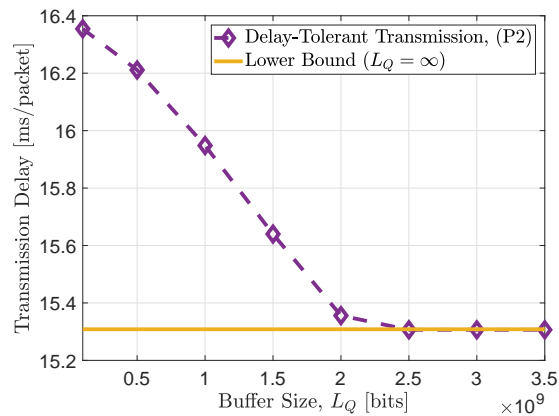}
    \caption{\tred{Transmission delay over buffer size $L_{\mathrm{Q}}$  [bits].}}
    \label{fig:Delay_BUF}
\end{figure}

Fig. \ref{fig:Throughput_BUF} shows the optimal average throughput of delay-tolerant transmission (P2) over the buffer size.
It is shown that the optimal average throughput is proportional to the given buffer size $L_{\mathrm{Q}}$.
With a larger buffer size given (i.e., being less constrained by the buffer constraint), the trajectory design of the UAV can be more flexibly optimized to maximize the average throughput of the mobile relay.
Note that the result of ``Delay-Tolerant Transmission with $L_{\mathrm{Q}}=\infty$" in Fig. \ref{fig:Throughput_BUF}, which considers the infinite buffer size, can be used as the upper bound of (P1).
As shown in Fig. \ref{fig:Throughput_BUF}, the average throughput results with $L_{\mathrm{Q}}\geq2.5 \cdot 10^{9}$ [bits] are close to the upper bound result.
Thus, we can verify that $L_{\mathrm{Q}}=2.5 \cdot 10^{9}$ [bits] is the optimal buffer size for this system.
Furthermore, it demonstrates that buffer size should be efficiently balanced to achieve better throughput performance.

\tred{
To show the relevance of delay and buffer size in the mobile relaying, we present transmission delay result in Fig. \ref{fig:Delay_BUF}.
This figure shows that transmission delay can be reduced with larger buffer size, and the transmission delay result converges when a sufficient buffer size (e.g., $L_{\mathrm{Q}}\geq2.5 \cdot 10^{9}$ [bits]) is given.
Based on Figs \ref{fig:Throughput_BUF} and \ref{fig:Delay_BUF}, we highlight that the large buffer size is beneficial for throughput performance as well as transmission delay.  }

\subsection{Simulation Results for Mixed FSO/RF Link Condition}
\tred{
To consider the atmospheric condition which is a crucial factor in mixed FSO/RF communication, we present the numerical result for different weather conditions.}

\begin{figure}[]
    \centering
    \includegraphics[width=0.48\textwidth]{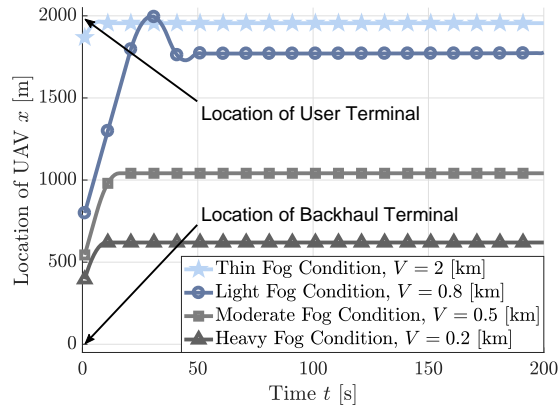}
    \caption{Trajectory comparison with optimized $x$-coordinate position over time $t$ with respect to different weather conditions.}
    \label{fig:X_V}
\end{figure}

\begin{figure}[]
    \centering
    \includegraphics[width=0.48\textwidth]{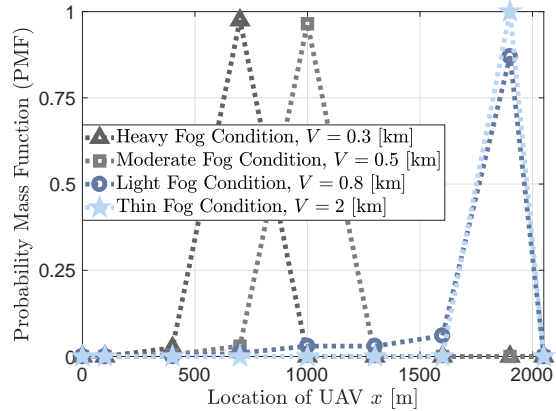}
    \caption{PMF based on $300$ [m] interval for Fig. \ref{fig:X_V}'s throughput maximized trajectories over different weather conditions.}
    \label{fig:Histogram_V}
\end{figure}

Figs. \ref{fig:X_V} and \ref{fig:Histogram_V} show the result of the delay-tolerant transmission case with infinite buffer size (i.e., $L_{\mathrm{Q}}=\infty$).
Fig. \ref{fig:X_V} represents the throughput maximized trajectories\footnote{Since the position $\mathbf{q}_{\mathcal{R}}[n]$, the velocity $\mathbf{v}_{\mathcal{R}}[n]$ and acceleration $\mathbf{a}_{\mathcal{R}}[n]$ of UAV are correlated by \eqref{C_q&v&a_P0}, we present the optimized trajectory $\mathbf{q}_{\mathcal{R}}[n]$ rather than $\mathbf{v}_{\mathcal{R}}[n]$ and $\mathbf{a}_{\mathcal{R}}[n]$. 
} in $x$-axis over different atmospheric conditions.
For thin fog condition (i.e., best atmospheric condition among the four weather conditions for FSO link), the UAV hovers near on the user terminal (e.g., $x=1955$ [m]) during whole time slots.
Whereas, for heavy fog condition, the UAV relay hovers near the backhaul terminal (e.g., $x=619$ [m]) to forward data. 
As the weather condition becomes worse, UAV tends to stay closer to the backhaul terminal to receive data without difficulty. 
Note that we consider the infinite buffer size to show the effect of weather conditions on the optimized trajectory of mobile UAV relaying.
Also, note that UAV transmits data to the user terminal via RF link even near the backhaul terminal.
Fig. \ref{fig:Histogram_V} shows PMF of each optimized trajectory in Fig. \ref{fig:X_V} based on $300$ [m] interval. 
As shown in Fig. \ref{fig:Histogram_V}, UAV hovers at a time of 96.5\% of the total flight time from $850$ [m] to $1150$ [m] in the moderate fog condition (e.g., $V=0.5$ [km]), while it hovers at a time of 87\% from $1750$ [m] to $2050$ [m] in the light fog condition (e.g., $V=0.8$ [km]).
It can be observed in Figs. \ref{fig:X_V} and \ref{fig:Histogram_V} that the better weather condition, the mobile relay hovers closer to the user terminal while easily receiving the data from the backhaul terminal. 
It is worth noting that, even in heavy fog condition (i.e., worst atmospheric condition among the weather conditions for FSO link), FSO can be considered as a valid option as a backhaul link. 
In addition, better weather conditions, such as clear condition, can yield the same result of trajectory and PMF of ``Thin Fog Condition $V=0.2$ [km]", while higher transmission rate of FSO link can be supported.

\subsection{Comparison with Conventional Scheme}

As a final remark, Table \ref{table:Comparison} shows the performance comparison of the conventional and proposed schemes.
This table compares the proposed schemes (e.g., (P1) and (P2)) with the baseline schemes (e.g., static relaying scheme and data-ferrying scheme \cite{Intro_1, Body_3, Num_1, Num_5}). 
\tred{
Static relaying (which is referred to as the fixed relaying system in \cite{Intro_1}) is a scheme where the UAV relay system stays in one position and transfers data.
Particularly, we consider that UAV (as a fixed relay) hovers at $x_{s}=1568$ [m] where the FSO and RF links have the nearly identical rates to maximize the achievable rate of the decode-and-forward relaying scheme (i.e., $R_{\mathrm{FSO}} \simeq R_{\mathrm{RF}}$) on the altitude of UAV $H=100$ [m], with a infinite buffer size $L_{\mathrm{Q}}=\infty$.
}
\tred{
We also consider another benchmark scheme called data-ferrying \cite{Num_1, Body_3}.
In this scheme, UAV first loads the data from $\mathcal{S}$ within some predetermined range $d_1$ from $\mathcal{S}$, flies towards $\mathcal{D}$ without any data reception or transmission and then de-queued the data to $\mathcal{D}$ when it is within the range $d_2$ from $\mathcal{D}$.
Specifically, the numerical results of the data-ferrying scheme yields from $d_1=d_2=300$ [m] and $L_{\mathrm{Q}}=\infty$.
}

In Table \ref{table:Comparison}, it can be found that the result of (P1) with the optimal buffer size $ L_{Q} = 2.5 \cdot 10^{9}$ [bits] and optimal delay limit $L^{\mathrm{req}}=5$ [slots] is tight enough to the result of (P2) which is the upper bound of the proposed delay-limited mobile relaying schemes.
In other words, even in delay-limited transmission, if appropriate $L^{\mathrm{req}}$ found, the optimal throughput for the system still can be achieved.
Note that the tighter delay requirement can be achieved at the expense of reduced average throughput.  
Even though there are additional buffer constraints and delay limit constraints, the proposed schemes achieve better throughput performance compared to the two baseline schemes, as shown in Table \ref{table:Comparison}.
Specifically, the resulting solution of (P1) with $L_{Q}=2.5 \cdot 10^{9}$ [bits] and $L^{\mathrm{req}}=5$ [slots] obtains about 223.33\% and 16.28\% gain compared to the static relaying scheme and the data-ferrying scheme, respectively.
\begin{table}[t]\footnotesize
	\centering
	\caption{\tred{Comparison of proposed scheme and conventional schemes.}}
	\resizebox{0.65\columnwidth}{!}	{\begin{tabular} { l | c }
			\toprule[1pt]
			\textbf{Relaying Schemes} & \textbf{Average Throughput {[}bps{]}} \\
			\midrule[0.75pt]
			Static Relaying \cite{Intro_1}      & \multirow{2}{*}{2.0203e+07} \\
			($x_s=1568$ [m], $L_{Q}=\infty$)    & \\
			\midrule[0.1pt]
			Data-Ferrying \cite{Num_1,Body_3}   & \multirow{2}{*}{5.6175e+07} \\
			($d_{1}, d_{2}=300$ [m], $L_{Q}=\infty$)    & \\
			\midrule[0.1pt]
			Delay-Limited Transmission Scheme (P1)    & \multirow{2}{*}{\textbf{6.5323e+07}} \\  
			($L_{Q}=2.5 \cdot 10^{9}$ [bits])            & \\
			\midrule[0.1pt]
			Delay-Tolerant Transmission Scheme (P2)   & \multirow{2}{*}{6.5332e+07} \\  
			($L_{Q}= 2.5 \cdot 10^{9}$ [bits])          & \\
			\bottomrule[1pt]
	\end{tabular}}
	\label{table:Comparison}
\end{table} 


\section{Conclusion} \label{Conclusion}

In this paper, we have studied the throughput maximization of mixed FSO/RF UAV-assisted mobile relaying with buffer and delay considerations.
For the optimization, we have designed an efficient trajectory for a UAV-enabled mobile relaying under the different weather conditions (e.g., attenuation conditions). 
In the mixed FSO/RF system with the transmission rate imbalance, the finite size buffer has been practically considered, and the effect of buffer size to the mobile relaying system has also been ascertained.
Furthermore, we have classified buffer constrained throughput maximization problem into two different transmission policies, i.e., delay-limited transmission and delay-tolerant schemes, to deal with the delay requirement.
To address these non-convex problems, we use the successive optimization algorithm.
Therefore, an efficient trajectory can be obtained by addressing convex QCP.
Through the numerical results, we validate the supremacy of the proposed algorithm over the conventional schemes and find the optimal buffer size and the optimal delay-time requirement. Moreover, we show the throughput-delay tradeoff for the system.




\bibliographystyle{IEEEtran} 
\bibliography{main}

\end{document}